\documentstyle[twoside,fleqn,espcrc2]{article}

\title{Blocking of Dynamical Triangulations with Matter}

\author{{\bf E.~Gregory}$^{\rm a}$, S.M.~Catterall$^{\rm a}$ and 
 G.~Thorleifsson\address{Physics Department, Syracuse University, 
 Syracuse, NY 13244, USA}}
       
\begin{document}

\begin{abstract}
We use the recently proposed node decimation algorithm for 
blocking dynamical geometries to investigate a class of models, 
with central charge greater than unity, coupled to $2D$ gravity. 
We demonstrate that the blocking preserves the fractal 
structure of the surfaces.
\end{abstract}

\maketitle

\section{MODEL}

The model we examine here is a two dimensional dynamically triangulated
surface coupled to Gaussian fields. It has the (fixed area)
partition function \cite{Boul86}
\begin{equation}
Z_N \;=\; \sum_{\{\tau,\phi\}} {\rm e}^{-S_\tau[\phi]},
\end{equation}
where the sum is over all combinatorial triangulations $\tau$ 
and field configurations $\phi$. The action,
\begin{equation}
S_\tau[\phi] \;=\; \sum_{\mu=1}^D \sum_{\langle i,j\rangle}
(\phi_i ^\mu -\phi_j^\mu)^2,
\end{equation}
depends on the configuration of Gaussian fields $\phi$.
Here $i$ and $j$ label nodes that are nearest neighbors.

This model is characterized by a central charge $c=D$, where $D$ is the 
number of Gaussian fields. For $1<c<25$ there exists no meaningful 
analytical solution. To test a recently proposed Monte Carlo 
renormalization group method, {\it node decimation} \cite{ThorCat96}, we 
investigated the cases $c=1$ and 
$c=10$.  The former was chosen because it is right at the limit of where 
a useful analytical solution can be found, and 
the latter because it was our expectation that a $c=10$ model should 
closely resemble the $c=\infty$ limit where the surface is known to be a 
branched polymer. 

\section{SIMULATIONS}

We update the lattice by proposing standard link flip moves, which are either 
accepted or rejected according to the outcome of a Metropolis test.  Similarly,
the field configuration is updated by subjecting a proposed change in the 
field at a node to a Metropolis test. In addition we have employed 
an overrelaxation update of the Gaussian fields, which we find to significantly
reduce autocorrelation times.

An overrelaxation move involves replacing the value of one of the fields as,
\begin{equation}
\phi_i \;\rightarrow\; \phi_i' \;=\; -\phi_i + \frac{2}{q_i}\sum_j\phi_j,
\end{equation}
where the nodes $j$ are the nearest neighbors of node $i$, and $q_i$ is its
coordination number.  We see that since 
\begin{equation}
\sum_j(\phi_i-\phi_j)^2 \;=\; \sum_j(\phi_i'-\phi_j)^2,
\end{equation}
the action is preserved automatically.  Hence, these moves are always 
accepted.    We find that using a ratio of only one Metropolis
field update to eleven overrelaxation updates decreases the autocorrelation 
time by an order of magnitude compared to pure Metropolis updates.

\section{MEASURING $\gamma_s$}

We chose to investigate how the value of the string susceptibility exponent 
$\gamma_s$ behaves under successive renormalization group transformations. 
The exponent $\gamma_s$ defines the behavior of the grand 
canonical partition function ${\cal Z}(\mu)$
near the critical value of the cosmological constant $\mu$: 
\begin{eqnarray}
{\cal Z}(\mu) &= &\sum_{N=0}^\infty Z_N e^{-\mu N} \\
 &\sim &(\mu - \mu_c)^{2-\gamma_s} \;\;\;\; 
 {\rm as} \;\;\; \mu \rightarrow \mu_c .
\end{eqnarray}

It turns out that the value of $\gamma_s$ is related to the size distribution 
of {\it minimal neck baby universes}. A minimal neck on a $2D$ dynamical 
triangulation is defined as a three link loop on the surface, that is not the 
boundary of one of the triangles.  Such a neck divides the
surface into two parts.  The baby universe is defined to be the smaller of 
the two parts.  It can be shown that for a surface of size $N$, the number of 
baby universes of size $B$ should go like \cite{Jain92}: 
\begin{equation}
\label{A}
 n_N(B) \;\sim\; B^{\gamma_s - 2}(N-B)^{\gamma_s - 2}.
\end{equation}
Therefore the recipe for finding $\gamma_s$ is as follows: First, search the 
surface for minimal necks. Second, upon finding one, count the number of 
triangles on each side of the surface and call the smaller of these two 
numbers $B$. Finally, fit the distribution $n(B)$, averaged over many 
surfaces, to Eq.~(\ref{A}) and extract $\gamma_s$. Since $\gamma_s$ 
characterizes the long distance properties of the surface, it is an 
appropriate observable with which to test how well the node decimation 
preserves the fractal structure of the surface.

\section{NODE DECIMATION}

\begin{table*}[hbt]
\setlength{\tabcolsep}{1.5pc}
\newlength{\digitwidth} \settowidth{\digitwidth}{\rm 0}
\catcode`?=\active \def?{\kern\digitwidth}
\caption[tab1]{The string susceptibility exponent $\gamma_s$ measured
 on surfaces obtained via node decimation (using $b=\sqrt 2$) }.
\label{tab:gammas}
\begin{tabular*}{\textwidth}{@{}l@{\extracolsep{\fill}}llllc} \hline
Blocking & $c=0$ & $c={1/2}$ & $c=1$ & $c=10$ \\  \hline
0 & -0.495(12) & -0.209(8)  & $-0.038(58)$ & 0.570(7) \\
1 & -0.498(7)  & -0.365(5)  & $\;0.030(57)$  & 0.562(16)\\
2 & -0.500(12) & -0.334(10) & $\;0.124(67)$  & 0.569(9)\\
3 & -0.495(15) & -0.341(6)  & $\;0.04(17)$   & 0.542(14)\\
4 & -0.505(19) & -0.315(11) &                & 0.582(11)\\
5 & -0.497(20) & -0.348(3)  &                & 0.493(21)\\
6 &            &            &                & 0.515(34)\\
\hline
Analytic result & $-{1/2}$ & $-{1/3}$ & 0 & ${1/2}$\\
\hline \hline
\end{tabular*}
\end{table*}

On a regular lattice implementing a real-space renormalization group is 
straightforward.  A familiar example is a field of Ising spins on a square 
lattice. Here one merely has to block the spins in some regular way,
for example by using a majority rule to convert each block of spins into one 
spin on the reduced lattice.  On a random lattice, such neat blocking is not
possible. Here the lattice itself is a dynamical 
variable. We need a way to reduce the size of the surface, while
retaining its large scale features. To accomplish this we require a Monte 
Carlo renormalization group adapted to random surfaces.

We have used the node decimation  
proposed in \cite{ThorCat96}. The main idea is to remove nodes, at random, 
to reduce the size of the surface.  It's obvious that removing a
node of coordination number greater that three will result in a 
non-triangular hole. To keep the lattice triangular throughout the
process, we first identify the node we want to remove, then flip links
around it until its coordination number is three. Only then do we remove the  
node, leaving a new triangle in its place. In rare cases, it may be impossible 
to reduce the coordination number of a node to three by flipping links, in
which case the change will be abandoned and the node restored to its original
condition.

The next step is to block the Gaussian fields themselves.  One 
obvious approach is to let the fields on the blocked nodes be given by
\begin{equation}
\label{B}
 \phi'_i \;=\; \xi \left [ (1-\alpha)\phi_i+\frac{\alpha}{n_i}\sum_j 
 \phi_j \right ].
\end{equation}
Here $n_i$ is the coordination number of the $i$-th node. The first term is
the direct contribution of the field at that node to the blocked
lattice, and the second term is an average of the neighbors of the node 
on the original lattice. We have introduced a relative weight $\alpha$
between those two terms, which in principle depends on the blocking 
factor $b=N/N'$, where $N$ and $N'$ are the volumes
of the bare and blocked lattices, respectively.  
Clearly $\alpha$ should go to zero as 
$b \rightarrow 1$. As $b$ becomes larger, $\alpha$ should grow as well. 
Choosing an $\alpha$ that is too large for a given blocking factor may,
on the other hand, result
in a cooling effect where the Gaussian fields on the blocked lattice are 
overly correlated.  Preliminary results indicate that this scheme is fairly 
robust under choice of $\alpha$. 

In general, we must re-scale the blocked fields by an overall factor $\xi$.
Simple arguments suggest that $\xi$ should be given by 
\begin{equation}
\xi \;=\; b^{\frac{-\beta}{d_H}},
\end{equation}
where $d_H$ is the fractal dimension of the surface and $\beta=-\eta/2$ is
the length dimension of the scalar field, defined by the scaling of the 
two-point function;
\begin{equation}
\label{*xx}
\langle \phi_i \phi_j \rangle \;\sim\; {r_{ij}^{\eta}}.
\end{equation}

From dimension counting we get the undressed 
length dimension $\beta_o$ for a scalar field in two dimensions;
\begin{equation}
H=\int d^2x(\partial \phi)^2  \;\;\;\Longrightarrow\;\;\; \beta_o=0.
\end{equation}
In the presence of quantum gravity $\beta_o$ must be replaced by $\beta$,
the dressed field dimension.  For $c \leq 1$ we can use the KPZ 
scaling \cite{KPZ}; 
\begin{equation}
\beta - \frac{\beta(1-\beta)}{1-\gamma_s} \;=\; \beta_o,
\end{equation}
where
\begin{equation}
 \gamma_s \;=\; {\textstyle \frac{1}{12}} 
 \left (c-1- \sqrt{(25-c)(1-c)} \right ).
\end{equation}
So, for $c=1$, $\beta = \beta_0 = 0$.  For $c>1$ however, 
the process is more complicated.
The scale factor $\xi$ can be determined numerically by measuring some 
operator ${\cal O}$, that has $n$ powers of $\phi$ in it.  
After $m$ blockings, ${\cal O}$ will scale like 
\begin{equation}
{\cal O}_m \;=\; \xi^{nm}{\cal O}_0.
\end{equation}

It is thus possible to determine $\xi$ by constructing a ratio of 
operators, obtained by different amount of blocking,
but compared at the {\it same} volume.  In particular, we have 
\begin{equation}
\xi^n \;=\; \frac{{\cal O}_{m+1}}{{\cal O}_m}.
\end{equation}
Preliminary investigations of the blocking of the Gaussian fields 
for $c=1$ seem to confirm that the overall scale factor $\xi=1$.

It is worth noting that the corresponding exponent $\eta$, has
recently been determined numerically for an Ising model
coupled to $2D$ gravity, by a direct fit to
Eq.~(\ref{*xx}) \cite{AmbAna96}.
 
\section{RESULTS}

We have measured the value of $\gamma_s$ as a function of blocking level 
for: $c=0$ (pure gravity), $c = \frac{1}{2}$ (Ising spins), 
$c = 1$ and $c = 10$.  Bare lattice $N_o=2000$ was used for $c=1/2$, 
$N_o=1000$ for the other models. These simulations were performed with 
several million sweeps apiece. The results 
are summarized in Table 1. 

These results clearly show that the node decimation preserves 
the analytically predicted value of $\gamma_s$, even 
after several iterations. In the case of an Ising model coupled to gravity, 
the measured value of $\gamma_s$ agrees more closely with the analytic 
prediction after one or two blocking iterations than on the bare lattice.
This indicates that the blocking procedure may be useful in minimizing finite 
volume effects.  The $c=1$ case is notoriously difficult to fit, and even after
fitting $n(B)$ to a form with logarithmic corrections the results have larger
errors.

\end{document}